# Optimizing defence, counter-defence and counter-counter defence in parasitic and trophic interactions – A modelling study

Stefan Schuster[1*], Jan Ewald[1,2], Thomas Dandekar[2], Sybille Dühring[1]

[1]Dept. of Bioinformatics, Friedrich Schiller University of Jena

Ernst-Abbe-Platz 2, 07743 Jena, Germany

[2]Dept. of Bioinformatics, Biocenter

Julius Maximilian University of Würzburg

Am Hubland, 97074 Würzburg

*Corresponding author, e-mail: stefan.schu@uni-jena.de

**Abstract**

In host-pathogen interactions, often the host (attacked organism) defends itself by some toxic compound and the parasite, in turn, responds by producing an enzyme that inactivates that compound. In some cases, the host can respond by producing an inhibitor of that enzyme, which can be considered as a counter-counter defence. An example is provided by cephalosporins, β-lactamases and clavulanic acid (an inhibitor of β-lactamases). Here, we tackle the question under which conditions it pays, during evolution, to establish a counter-counter defence rather than to intensify or widen the defence mechanisms. We establish a mathematical model describing this phenomenon, based on enzyme kinetics for competitive inhibition. We use an objective function based on Haber's rule, which says that the toxic effect is proportional to the time integral of toxin concentration. The optimal allocation of defence and counter-counter defence can be calculated in an analytical way despite the nonlinearity in the underlying differential equation. The calculation provides a threshold value for the dissociation constant of the inhibitor. Only if the inhibition constant is below that threshold, that is, in the case of strong binding of the inhibitor, it pays to have a counter-counter defence. This theoretical prediction accounts for the observation that not for all defence mechanisms, a counter-counter defence exists. Our results should be of interest for computing optimal mixtures of β-lactam antibiotics and β-lactamase inhibitors such as sulbactam, as well as for plant-herbivore and other molecular-ecological interactions and to fight antibiotic resistance in general.

## 1. Introduction

In many parasitic and trophic interactions, the host or prey organisms, respectively, defend themselves by toxic or malodorous compounds (defence chemicals) or other mechanisms such as pH changes, fever etc. Many of the attacking organisms, in turn, produce enzymes degrading the defence chemicals, which can be considered as a counter-defence (in the case of parasites often called evasion mechanisms). An example is the defence of human immune cells against the pathogenic fungus *Candida albicans* by producing superoxide radicals and other reactive oxygen species (ROS). *C. albicans* and many other pathogens, in turn, can detoxify the ROS by superoxide dismutases (SODs) (Frohner et al., 2008). For an overview of host-pathogen interactions between the human innate immune system and *C. albicans*, see Dühring et al. (2015).

The question arises whether the host or prey can inhibit the evasion mechanisms by counter-counter defences. Does it pay, during evolution, to establish a counter-counter defence or is it more favourable to intensify or widen the defence mechanisms? Analogously, does it pay for the parasite or pathogen to establish a counter-defence or is it more favourable to intensify or widen the attack mechanisms? Are there examples of even higher levels of back-and-forth interactions?

Such questions can be tackled in the framework of optimization approaches. Optimality principles have widely been used in explaining mechanisms and properties of living cells and organisms (cf. Heinrich and Schuster, 1996; Cornish-Bowden, 2016). For example, the optimal half-saturation constant of haemoglobin has been calculated by assuming that oxygen is bound and released in an optimal way (Willford et al., 1982). Waddell et al. (1999) could explain the ATP stoichiometry of glycolysis based on the principle of maximum ATP production rate (for other ATP-producing pathways, see Werner et al., 2010). To analyse the usage of defence mechanisms in a mathematical way, the optimal defence hypothesis is helpful: It states that organisms allocate their defences in a way that maximises fitness (cf. Radhika et al., 2008). In the context of host-pathogen interactions, also dynamic (time-dependent) optimization was used (Dühring et al., 2017).

In this paper, we use mathematical modelling and enzyme kinetics to study the benefit of counter-counter defences. We first review several examples of counter-counter defences in various kingdoms of life (Section 2). In Section 3 and Appendix A, we present our mathematical model and analyse the effect of a substance inhibiting an enzyme that degrades a defence chemical. We compare this with an intensified defence. In that computation, we use analytical (rather than purely numerical) calculations as far as possible.

## 2. Examples of counter defences and higher-order counter defences

### 2.1. Fungi or Streptomyces and bacteria

The constellation considered in this paper is illustrated in a general way in Fig. 1A and an example is shown in Fig. 1B.

A)

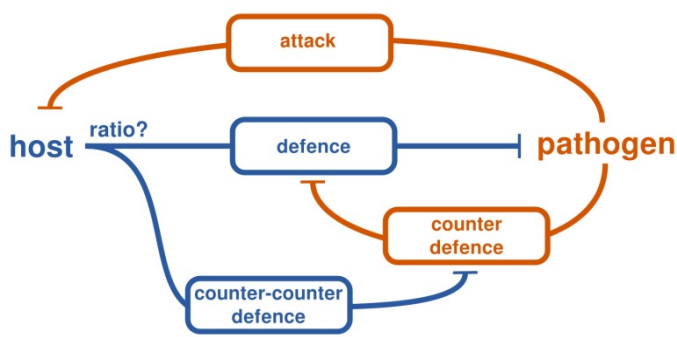

B)

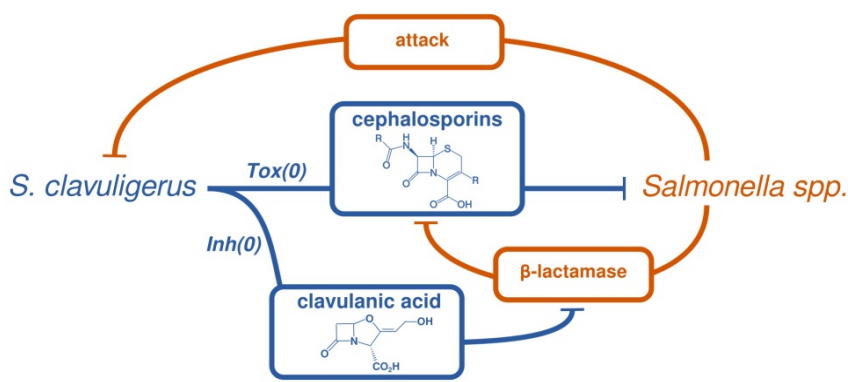

**Fig. 1. Scheme of molecular host-pathogen interactions.** A) General scheme including four levels of interaction (not all of them being always relevant). A similar constellation can also be found in predator-prey or plant-herbivore interactions. B) Specific example of *Streptomyces clavuligerus* and *Salmonella* bacteria interacting via β-lactam antibiotics, β-lactamase and clavulanic acid.

An illustrative example is the following: Antibiotics represent a defence of certain fungi against attacks or competition for nutrients by bacteria. A special class is formed by the β-lactam antibiotics, which are called that way because they harbour a β-lactam ring formed from cysteine and valine (Figure 1B). An example is provided by penicillins, such as those produced by *Penicillium chrysogenum* (cf. Christensen et al., 1995) and *Aspergillus nidulans* (Brakhage et al., 1994). Further groups of β-lactam antibiotics are the cephalosporins produced by fungi of the genus *Acremonium*, formerly known as *Cephalosporium* (Demain and Zhang, 1998) and the bacterium *Streptomyces clavuligerus* (Aharonowitz and Demain, 1978) as well as the synthetic carbapenems and semi-synthetic monobactams (Drawz and Bonomo, 2010). In medical applications, also semi-synthetic penicillin derivatives such as ampicillin and amoxicillin are widely used.

As a counter-defence, many gram-negative bacteria, for example, *Salmonella spp.* and *Klebsiella pneumoniae* respond by secreting enzymes called β-lactamases. These enzymes are able to cleave the β-lactam ring, which makes these antibiotics inefficient because an essential mode of their action is that the ring opens at the target site. There are a wide variety of different β-lactamases such as metallo-β-lactamases (cf. Paterson et al., 2003).

*S. clavuligerus*, in turn, produces clavulanic acid, which is structurally similar to the core unit of penicillins, with the sulfur atom replaced by oxygen (Figure 1B). In particular, it involves a β-lactam ring and can, therefore, act as a competitive inhibitor of β-lactamases. It shows little antimicrobial activity alone (Drawz and Bonomo, 2010). Rather, it provides a counter-counter defence. There are also synthetic pharmaceutical drugs that inhibit β-lactamases (Mezes et al., 1982), for example, those distributed under the trade names sulbactam and tazobactam (Noguchi and Gill, 1988). For the above-mentioned reasons, these are often administered in combination with β-lactam antibiotics.

The synthetic β-lactam antibiotic, carbapenem, is largely insusceptible to the classical β-lactamases. However, it can be degraded by metallo-β-lactamases. These, in turn, are inhibited by aspergillomarasmine A produced by *Aspergillus versicolor* (King et al., 2014). It is unclear, though, why that fungus produces that counter-counter defence molecule because, to our knowledge, it has not been found that the fungus would produce β-lactam antibiotics. It may be hypothesized that it is useful in microbial communities involving producers of such antibiotics.

*2.2. Plants and herbivores*

An example from the realm of plant-herbivore interactions is the following: The South-American plant *Psychotria viridis* produces the hallucinogenic compound dimethyltryptamine as a defence chemical (cf. Soares et al., 2017). Higher animals produce, mainly in the brain, monoamine oxidase, which degrades tryptamines. *P. viridis*, in turn, also produces beta-carbolines (a.k.a. harmane compounds). These inhibit monoamine oxidase and, thus, represent a counter-counter defence.A more sophisticated example is the following: Glucosinolates are activated in Brassicaceae by the enzyme thioglucoside glucohydrolase (myrosinase). That activation leads to the toxic compound isothiocyanate. Some insects can degrade glucosinolates to less toxic nitriles. When the plants sense this degradation by insects, they form the nitriles themselves (Mumm et al., 2008). This is paradoxical at first sight, but these substances prevent insect females from oviposition on the plants because they "assume" that there is a caterpillar already. Thus, the insects erroneously try to avoid intraspecific competition. This is a true counter-counter defence because it would not be effective if the counter-defence (degradation to nitriles by insects) were not present.

*2.3. Plants and pathogens*

Upon contact with pathogens, many plants (e.g. *Arabidopsis*) suppress the opening of their stomata by special mitogen-activated protein kinases (MAPK), e.g. MPK3 and MPK6. This is an example of plant immunity. The bacterial pathogen *Pseudomonas syringae* produces coronatine, an analogue of jasmonic acid, which suppresses these MAPKs. The counter-counter defence of the plant consists in the suppression of the jasmonate pathway by specific phosphatases, so that the reopening of stomata is suppressed (Mine et al., 2017).

There are further examples concerning the interaction of plants and pathogens, where at least one stage of interaction resides at the transcript level or genetic level (Pumplin and Voinnet, 2013, Hutin et al., 2015). For example, plant-pathogenic *Xanthomonas* bacteria use Transcription Activator-Like (TAL) effectors belonging to their type three secretion system (T3SS), which is an attack mechanism (Hutin et al., 2015). The TAL effectors bind to a programmable DNA-binding domain and induce specific plant genes, which suppress plant immunity and facilitate nutrient acquisition by the bacteria. Plants, in turn, have evolved defence strategies against the action of TAL effectors. One mechanism is to avoid binding by mutations, resulting in loss of susceptibility. Bacteria, in turn, may overcome the effect of mismatches between TAL effectors and their cognate binding elements by aberrant-length repeats that can "loop out" when the TAL effector binds. This allows the recognition of target DNA sequences with a −1 nucleotide frame-shift.

Hutin et al. (2015) speak of several rounds of interaction: the first to fourth rounds corresponding to attack, defence, counter defence and counter-counter defence, respectively. For the plant-bacteria example, the fourth round has not yet been observed in nature, but can be realized by human intervention. In view of application, this scenario is of great interest. As plant immunity is a defence mechanism, the TAL effectors (which suppress immunity) could even be considered both as an attack mechanism and as a counter defence. According to that interpretation, the third round described above would even be a $(counter)^3$ defence.

*2.4. Macrophages and pathogens*

The levels of defence are not always clear-cut. Counter-counter defences may overlap with direct defences. For example, the Pra1 protein (counter-defence) of *C. albicans* interacts with human factor H, which is bound by *C. albicans* as an evasion mechanism (Zipfel et al., 2011). In human phagolysosomes, the pH is lowered considerably, which inhibits Pra1. However, lowering the pH is a broad defence against many proteins of *C. albicans* (Mayer et al., 2013; cf. Dühring et al., 2015). Interestingly, *C. albicans* and other fungal pathogens are able to increase the pH value from ~4 to ~7.5 by generation and export of ammonia (Vylkova, 2017). If acidification by the host is considered as a counter-counter defence, alkalinisation by the fungus would even be a $(counter)^3$ defence.

Another striking example of multi-layer defence interactions was discovered in the interaction between macrophages and bacterial pathogens like *Yersinia pestis* or *Pseudomonas aeruginosa*. As a first line of defence, macrophages phagocytose the attacking bacteria and, among other antibacterial strategies, are limiting glucose and other carbon sources within the phagolysosome (Weiss and Schaible, 2015). A key factor for pathogenicity of many bacteria such as *Y. pestis* is their ability to survive within macrophages by utilizing acetate and fatty acids (Li and Yang, 2008; Fukuto et al., 2010). This is enabled by the enzymes of the glyoxylate shunt, which circumvent the decarboxylation steps of the TCA cycle and are absent from most animals (Dolan and Welch, 2018). Michelucci et al. (2013) found that macrophages produce itaconic acid, which acts as a competitive inhibitor of the isocitrate lyase catalyzing the first reaction of the glyoxylate shunt. However, this counter-counter defence by macrophages is abolished in bacterial pathogens like *Y. pestis* by a recently discovered degradation pathway of itaconate to pyruvate (Sasikaran et al., 2014), which represents a (counter)³ defence. Interestingly, the genes of the degradation pathway preferably occur in pathogens. This shows that the depth of defence interactions is certainly optimized by each organism depending on its survival strategy.

## 3. Mathematical modelling: Defence chemical, degrading enzyme and inhibitor of the enzyme

*3.1. Basic differential equation*

We consider a defence chemical (toxin, e.g. penicillin) and an enzyme degrading that toxin (e.g. β-lactamase). We want to answer the question what is more beneficial for the toxin producer: producing more toxin or producing, in parallel, some inhibitor of the degrading enzyme (Fig. 1).

To describe the rate of degradation of the toxin, we use the most widely used enzyme kinetics for competitive inhibition (cf. Cornish-Bowden, 2012):

$$\frac{\mathrm{d}Tox}{\mathrm{d}t} = -\frac{V_{\max} Tox}{K_\mathrm{M}(1+Inh/K_\mathrm{I})+Tox} \tag{1}$$

with the following symbols: $V_{\max}$ and $K_\mathrm{M}$, maximal velocity and Michaelis constant of the enzyme, respectively; *Tox* and *Inh*, concentrations of toxin and inhibitor, respectively; $K_\mathrm{I}$, inhibition constant. Note that $K_\mathrm{I}$ is a dissociation constant. Thus, the lower $K_\mathrm{I}$ is, the more efficient the inhibitor will be.

Importantly, Eq. (1) only describes the effect of competitive inhibitors if these act reversibly. Some inhibitors of β-lactamases such as avibactam (Ehmann et al., 2012; Lahiri, 2014) and boronic acid transition state analogues (Drawz and Bonomo, 2010) are indeed reversible, while clavulanic acid, sulbactam and tazobactam are irreversible inhibitors (Drawz and Bonomo, 2010). The former do not harbour a β-lactam ring. The latter bind, in a first phase, in a reversible way, so that the initial enzyme velocity can be described by Eq. (1). Thereafter, the enzyme-inhibitor complex is converted in a first-order reaction to an inactive form (so-called suicide inhibition). This conversion is usually described by an exponential time function (Labia et al., 1980).

In the present paper, we consider reversible inhibition. As an approximation, Eq. (1) can even be used for irreversible inhibitors if inactivation proceeds very slowly. For example, for the β-lactam sulfon CP 45899 (being an irreversible β-lactamase inhibitor), the rate of degradation is about 1000 times lower than that of penicillin degradation (Labia et al., 1988) and can, thus, approximately be neglected.

It is a plausible assumption that the toxin producer has a certain capacity $C$ to invest into the production of toxin and/or inhibitor:

$$Tox(0) + Inh = C. \tag{2}$$

The capacity $C$ expresses in which amount the fungus can produce penicillin or structural analogues, measured by the concentration in which they can be maintained in the vicinity of the fungus. For simplicity's sake, we assume that the synthesis of the toxin and its analogues imply the same costs per molecule to the producer. This assumption is justified in the case of a structurally similar competitive inhibitor. For example, penicillin and clavulanic acid are so similar in structure that their synthesis can be assumed to imply comparable costs. However, it is worth mentioning that these two substances are synthesized on two different pathways, with arginine being a precursor of clavulanic acid (Townsend, 2002). In the case of different costs, Eq. (2) can easily be modified by a weighting coefficient assigned to $Inh$.

Eqs. (1) and (2) can be combined into:

$$\frac{dTox}{dt} = -\frac{V_{max}Tox}{K_M(1+(C-Tox(0))/K_I)+Tox} \tag{3}$$

In some of the following equations, we use the effective Michaelis constant

$$K'_M = K_M\left(1 + \frac{C-Tox(0)}{K_I}\right). \tag{4}$$

While the inhibitor level remains constant in time, the toxin concentration decreases due to degradation (Fig. 2). In Eq. (3), the toxin concentration at time zero is relevant.

*3.2 The optimality criterion*

We now tackle the question what the optimal allocation to toxin and inhibitor is. If no inhibitor was produced, the initial toxin concentration would be maximum. However, also the degradation rate would be maximum. This is visualized in Fig. 2. The other extreme situation is where the entire capacity would be invested into the production of inhibitor. This is certainly useless because no toxic effect against the counterpart would remain.

We quantify the defence effect of the toxin by the time integral over *Tox(t)* from zero to infinity. In toxicology, this is known as Haber's rule (cf. Connell et al., 2016). In pharmacokinetics, that time integral is called Area under the curve (AUC) (cf. Wagner et al., 1985; Lappin et al., 2006).

We write the optimization principle as

$$\text{maximize } \int_0^\infty Tox(t)\, dt, \tag{5}$$

where the initial value *Tox*(0) is the variable of optimization.

The AUC is finite only when *Tox(t)* is a monotonic decreasing function approaching zero and decreases fast enough. For example, the time integral over an exponentially decreasing function is finite, while the time integral over a hyperbolic function 1/*t* is infinite. Integrating up to infinite times is a mathematical abstraction. In reality, the effect ends at finite times, of course. However, the 'tail' of the integral can often be neglected.

## 4. Results

### 4.1. Solution by linear approximation

In some equations (especially longer ones), we will use the abbreviations $Tox(t) = T$ and $Tox(0) = T_0$. To get a preliminary idea of the solution, we first derive an approximative solution by linearization and below, we solve the nonlinear problem. For the approximation, we assume that *T* is considerably lower than the Michaelis constant. This is particularly well justified for long times (note that the integral (5) spans infinite times), because then the β-lactamases degrade the penicillin to low levels. Under this assumption, we can simplify the Michaelis-Menten kinetics to linear kinetics:

$$\frac{dT}{dt} = -\frac{k*T}{1+(C-T_0)/K_\mathrm{I}}, \tag{6}$$

where we abbreviate the ratio $V_\mathrm{max}/K_\mathrm{M}$ by *k*. The time course of the penicillin concentration then decreases in an exponential way:

$$T = T_0 * exp\left(-\frac{k}{1+(C-T_0)/K_\mathrm{I}}\right)t. \tag{7}$$

The time course is illustrated in Fig. 2.

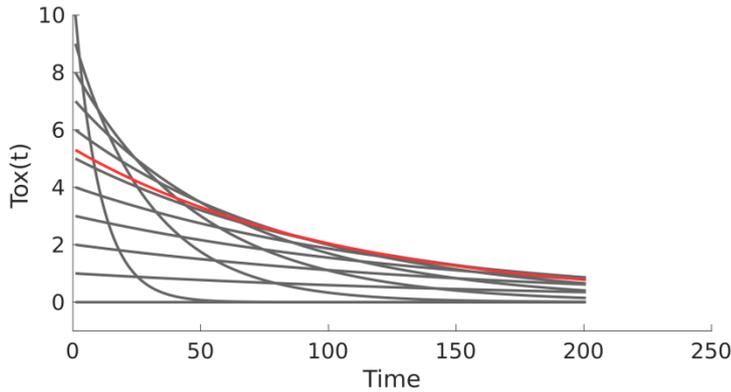

**Fig. 2. Time course of toxin concentration without inhibitor and with varying concentrations of the inhibitor of the enzyme degrading the toxin.** Curves were computed by approximation (3), which leads to exponential function (4). The curves correspond to increasing inhibitor concentrations, from the curve starting with the highest initial value corresponding to the case without inhibitor up to the lowest curve, where $I = C = 10$, with a step size of 1. Red curve: optimal case. For the bottom curve, the toxin concentration is zero all the time because all the capacity is invested into the inhibitor. Parameter values: $k = 0.1$; $C = 10$; $K_i = 0.5$.

The integral can be calculated analytically:

$$\int_0^\infty T \, dt = T_0 \int_0^\infty exp\left(-\frac{k}{1+(C-T_0)/K_I}\right) t \, dt$$

$$= \frac{T_0}{k}(1 + (C - T_0)/K_I) \,. \tag{8}$$

We can multiply this expression by $K_I * k$ without changing the position of the maximum, so that the maximization principle reads

$$\text{maximize } T_0(K_I + C - T_0) \,. \tag{9}$$

This is a quadratic function in $T_0$ (Fig. 3) and, thus, has indeed a maximum, which can be calculated by differentiation. This gives

$$T_{0,opt} = \frac{K_I + C}{2} \,. \tag{10}$$

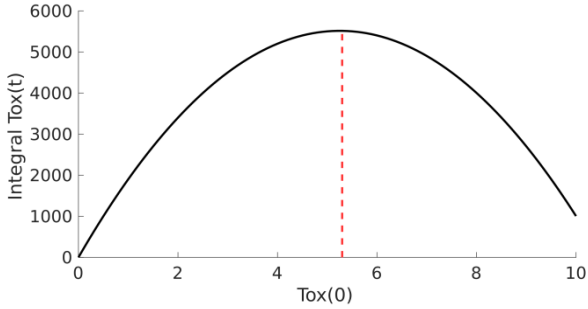

**Fig. 3. Integrated toxic effect (objective function value) vs. initial toxin concentration, *Tox*(0), as computed by the approximative function (9).** Note that $Tox(0) = C-Inh$. Dashed red line, maximum at $Tox(0) = (K_I+C)/2 = 5.25$. Parameter values are as in Fig. 2.

Now we can distinguish two cases:

(i) $K_I < C$ (strong binding of the inhibitor). Then the benefit-to-cost ratio is high because producing the inhibitor provides a high benefit. The maximum is in the interior of the interval [0, *C*]. It pays to invest part of *C* into the counter-counter defence (e.g. clavulanic acid). In the extreme case where $K_I$ is nearly zero (very efficient inhibitor), about half of *C* should be invested into the defence and half into the counter-counter defence.

(ii) $K_I > C$ (weak binding of the inhibitor, or low benefit-to-cost ratio): Then the maximum lies above *C* and, thus, cannot be reached. It does not pay to invest part of *C* into the counter-counter defence. It is better to invest *C* fully into toxin (e.g. penicillin). This is because the counter-counter defence is too weak.

*4.2. Exact analytical solution*

Eq. (1) cannot be solved analytically in a way that *Tox* could be written as a function of time. Nevertheless, the improper integral occurring in Eq. (5) can be calculated analytically for the case where the substance (toxin in our case) is degraded by a Michaelis-Menten enzyme (cf. Wagner et al., 1985). To that end, Eq. (6) is rearranged as

$$Tox \mathrm{d}t = -\frac{K'_M}{V_{max}} \mathrm{d}Tox - \frac{Tox}{V_{max}} \mathrm{d}Tox \qquad (11)$$

where the abbreviation (4) is used. Integration gives

$$\int_0^\infty Tox \mathrm{d}t = -\frac{K'_M Tox}{V_{max}} \Big]_{T_0}^0 - \frac{Tox^2}{2V_{max}} \Big]_{T_0}^0$$

$$= \frac{1}{V_{\max}} \left( K'_M T_0 + \frac{T_0^2}{2} \right). \tag{12}$$

Note that the lower and upper bounds in the integrals over *Tox* are $T_0$ and 0, respectively, because the curve proceeds between these two values. Like the function in Eq. (9), the function in Eq. (12) is quadratic in $T_0$ (Fig. 5). It is consistent with the objective function (9) in the approximative solution, as can be seen by assuming $T_0 \ll K_M$, which implies $T_0 \ll K'_M$. Thus, we can neglect, in that case, the term $T_0^2/2$ in the parenthesis in Eq. (12). Under consideration of the abbreviation (4), this leads to an objective function that is proportional (and, thus, equivalent) to the one in Eq. (9). An alternative calculation of the improper integral is presented in the Appendix.

Fig. 4 shows the resulting time course. The curves starting at higher toxin concentrations (that is, for which the inhibitor concentration is lower) decrease, at the outset, almost with a constant slope, unlike in the approximative solution. This is because the parameter values were chosen such that the enzyme operates near to saturation first, so that the degradation velocity is nearly constant. This can also be seen from Eq. (A3) in the Appendix. Since the logarithmic function increases less than the linear function, the term $T$ dominates the term $\ln T$ for large $T$. Thus, $T$ and time are then correlated nearly linearly. Later (for larger time values), when the concentration has become so low that the enzyme operates in the linear range, it decreases in an exponential way, like in the approximative solution.

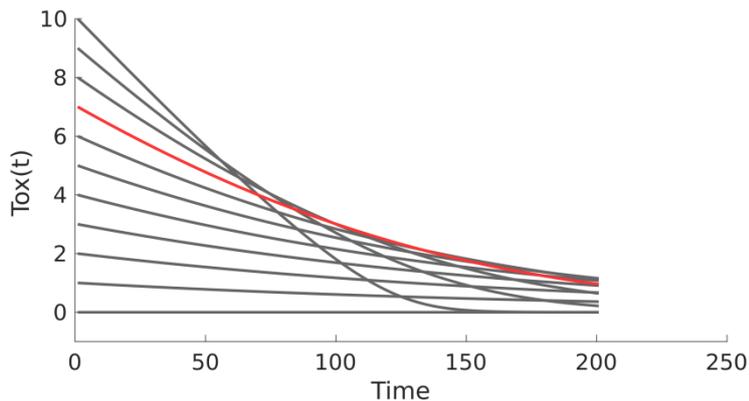

**Fig. 4. Time course of toxin concentration for the exact solution.** The time course can be computed analytically by Eq. (A3) in the Appendix or numerically using the kinetics for competitive inhibition (1). The curves correspond to increasing inhibitor concentrations, from the curve starting with the highest initial value corresponding to the case without inhibitor up to the lowest curve, where $I = C = 10$, with a step size of 1. Red curve: optimal case. Parameter values: $V_{\max} = 0.1$; $K_m = 1$; $C = 10$; $K_i = 0.5$.

Returning now to the general case of not necessarily small $T_0$ values, we calculate the $T_0$ for which the function in Eq. (9) is maximum. By differentiation of the r.h.s. of Eq. (12), we obtain the position of the maximum:

$$T_{0,opt} = \frac{K_M(K_I+C)}{2K_M-K_I}. \tag{13}$$

It can be seen that there is no optimum at a positive $T_0$ value if $K_I > 2K_M$, in which case the inhibitor is very weak.

By comparing the optimal $T_0$ value with $C$, we can consider three different cases:

(i) $K_I < K_M C/(K_M + C)$ (strong binding of the inhibitor). The benefit-to-cost ratio is high and the maximum lies in the interior of the interval [0, $C$] (Fig. 5). It pays to invest part of $C$ into the counter-counter defence. Like in the linearized case, in the limit $K_I \to 0$, the optimal $T_0$ tends to $C/2$, implying that about half of $C$ should be invested into the defence and half into the counter-counter defence.

(ii) $2K_M > K_I > K_M C/(K_M + C)$ (weak binding of the inhibitor, or low benefit-to-cost ratio): Then the maximum lies above $C$ and, thus, cannot be reached.

(iii) $K_I > 2K_M$. There is no optimum at a positive $T_0$ value. Note that this case does not occur in the linearized case because the kinetics can only be linearized when the $K_M$ value is high.

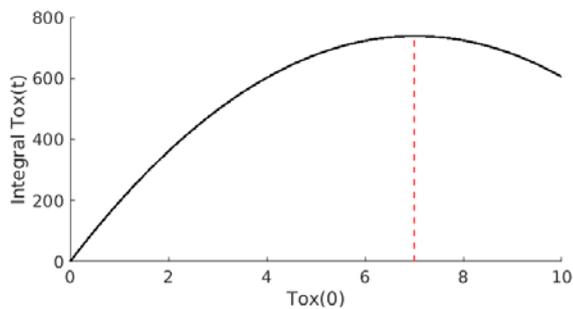

**Fig. 5**. **Integrated toxic effect (objective function value) vs. initial toxin concentration, *Tox*(0), for the exact solution (12).** Note that *Tox*(0) = *C−Inh*. Dashed red line, maximum *Tox*(0) = 7.0. Parameter values as in Fig. 4.

Both in cases (ii) and (iii), it is better to invest $C$ fully into toxin (e.g. penicillin) than to invest part of $C$ into the counter-counter defence. This is because the counter-counter defence is too weak.

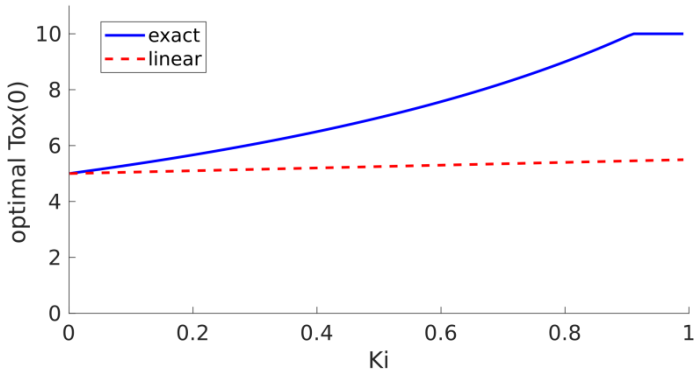

**Fig. 6. Comparison of results for the optimal initial toxin concentration between the approximative and exact solutions.** Dashed red line, linearized model; blue solid curve, nonlinear model. Parameter values: $V_{max} = 0.1$; $K_m = 1$; $C = 10$.

Fig. 6 shows a comparison between the exact and linearized solutions. It can be seen that the exact curve hits the upper limit value of 10 at a $K_I$ value of about 0.9. This can be calculated by $K_M C/(K_M + C) = 10/(1 + 10) = 0.909$. For $K_I$ values larger than this threshold, the optimal $Tox(0)$ value equals $C$. The curve corresponding to the linearized case hits the upper limit value at $K_I = 10$. It turns out that the linearization yields a good approximation only for very low $K_I$ values. This is because $K_M$ is quite low; for larger values the approximation gets better (not shown).

## 5. Discussion

In this paper, we have analysed, by mathematical modelling, the defence, counter-defence and counter-counter defence in the interaction between fungal pathogens and the human host. As an illustrative example, we have studied β-lactam antibiotics (defence), β-lactamases (counter defence) and clavulanic acid (counter-counter defence). The calculations can be applied to any interaction among two opposing organisms, in which a toxin is degraded by an enzyme that is, in turn, inactivated by a competitive inhibitor.

The optimal allocation can be calculated in an analytical way in spite of the nonlinearity of the underlying differential equation. The calculation provides a threshold value for the inhibition constant: $K_M C/(K_M + C)$, where $K_M$ is the Michaelis constant of the enzyme and $C$ denotes the total capacity that can be invested by the host in the toxin and the inhibitor. Only if the inhibition constant is below this threshold, that is, in the case of strong binding of the inhibitor, it pays to have a counter-counter defence.

This theoretical prediction accounts for the observation that not for all defence mechanisms, a counter-counter defence exists. It can be assumed that host organisms that do not exhibit such mechanisms are not capable of producing inhibitors that would be efficient enough to fulfil the above-mentioned criterion. This explains well why some fungi do produce β-lactamase inhibitors while others do not.

Our results are likely to have very useful applications in pharmacy. They allow one to compute the optimal mixture in the dosage of β-lactam antibiotics and β-lactamase inhibitor such as sulbactam and tazobactam in combination pharmaceuticals.

It is interesting that the total effect of a toxin in the case of permanent degradation by Michaelis-Menten type enzyme, as calculated according to Haber's rule, can be calculated analytically (Wagner et al., 1985). In the Supplement, we have presented an alternative analytical calculation, which is based on the idea to swap variables, as used already by Michaelis and Menten (1913).

The use of clavulanic acid, which has a similar structure as the β-lactam antibiotics, by *Streptomyces clavuligerus* is interesting also in view of the question why many defence chemicals (such as glucosinolates in Brassicaceae plants) occur as mixtures of several similar compounds. One might argue that mixtures widen the effect. On the other hand, the counterpart organism could respond by degrading enzymes with broad substrate specificity in a way that all the defence chemicals in the mixture are inactivated. However, the advantage of mixtures may be that some chemicals may act as competitive inhibitors of those enzymes. Moreover, broad substrate specificity often implies larger $K_m$ values, so that it is easier to synthesize an inhibitor that corresponds to case (i) in Subsection 4.2.

The results are of general relevance. They show that it depends on certain conditions whether a counter-counter defence is useful or whether it is better to intensify the defence. It is plausible to assume that the results can be generalized in a qualitative way to cases of counter-counter defence other than inhibitors of degrading enzymes. As mentioned in the Introduction, also the interplay at the level of attack and counter defence can be studied in a similar way. Our results suggest that it is only worth establishing a counter-counter defence (or counter defence) rather than intensifying an existing defence (respectively attack) if the new level is more effective than a critical value. In the case of enzyme inhibitors, effectiveness can be measured by the inhibition constant.

The presented model can be a guideline in searching for hitherto unknown effectors, for example, inhibitors of superoxide dismutases (SODs) or secreted aspartyl proteases (SAPs). However, since the substrate of SODs are reactive oxygen species (ROS), the question arises whether competitive inhibitors for that exist.

In future extensions of the study, it is worth including the host microbiome. It may well be that some bacteria produce antifungal substances that can, from the host's viewpoint, be considered as a defence or counter-counter defence. Further, this extends the constellation of host-pathogen interactions (Fig. 1) to more than two organisms. This enables an investigation of communities where organisms show a division of labour between defence and counter-counter defence mechanisms against a pathogen or competitor.

A more sophisticated model should include the reliability of the effector. As each level depends on the lower levels, an effector becomes useless if the effector on the next-lower level is qualitatively changed. This may shift the trend away from establishing new levels towards intensifying existing effects on lower levels.

As the present model is based on Haber's rule, the objective function also involves contributions from time periods in which the toxin concentration is low. A more realistic approach should consider that there is a critical threshold of the toxin concentration below which the defence chemical has no effect at all. Moreover, it is worth analysing the effect not only of a change in the inhibition constant but also in the capacity $C$. For example, counter-counter defence are induced upon confrontation with the counterpart, which may increase the capacity allocated to the defence.

**Ackowledgments**: The authors acknowledge support by the Deutsche Forschungsgemein-schaft (DFG) within the CRC/Transregio 124 'Pathogenic fungi and their human host' (project number 210879364), subproject B1. We thank Jan Grau, Daniel Schober (both Halle), Jonathan Gershenzon (Jena) and Kenichi Tsuda (Cologne) for drawing our attention to several examples of counter-counter defence and Axel Brakhage, Falk Hillmann, Hortense Slevogt (all in Jena), and Barbara Bakker (Groningen) for stimulating discussions.

**Appendix: Alternative analytical solution**

As has been shown already by Michaelis and Menten (1913) in their pioneering paper on enzyme kinetics, such equations can be solved by exchanging the dependent and independent variables, that is, in our case, *Tox* and time. We obtain

$$\int_{T_0}^{T} \frac{K_M' + T}{T} dT = -\int_0^t V_{max} dt \ . \tag{A1}$$

$$K_M' \ln \frac{T}{T_0} + T - T_0 = -V_{max} t. \tag{A2}$$

$$t = \frac{1}{V_{max}}(K'_M(\ln T_0 - \ln T) + T_0 - T). \tag{A3}$$

Thus, we can formally express time as a function of the toxin concentration. Although this function cannot be inverted analytically, it can be used to plot *Tox* as a function of time, by swapping coordinate axes (Fig. 4). An alternative way of computing this function is by integrating the differential equation (1) numerically.

For computing the integral (5), we make use of the fact that the area under the curve $T(t)$ is the same as the area under the curve $t(T)$. Therefore, we can calculate the integral $\int_0^{T_0} t dT$ using Eq. (A3).

$$\int_0^{T_0} t \, dT = \frac{1}{V_{max}} \left\{ K'_M T(\ln T_0 - \ln T + 1) + T_0 T - \frac{T^2}{2} \right]_0^{T_0} \right\}$$

$$= \frac{T_0(2K'_M + T_0)}{2V_{max}}$$

$$= \frac{T_0\left(2K_M\left(1+\frac{C-T_0}{K_I}\right)+T_0\right)}{2V_{max}}. \tag{A4}$$

This resulting term is equivalent to that obtained in Eq. (12).